\documentclass[twocolumn,superscriptaddress,prl,floatfix,showpacs]{revtex4}
\usepackage{graphicx,bm}
\begin{document}

\title{Supernova Inelastic Neutrino-Nucleus Cross Sections from
High-Resolution Electron Scattering Experiments and Shell-Model
Calculations}
\author{K. Langanke}
\affiliation{Institute for Physics and Astronomy,
University of {\AA}rhus,
  DK-8000 {\AA}rhus C, Denmark}
\author{G. Mart\'{\i}nez-Pinedo}
\affiliation{ICREA and Institut d'Estudis Espacials de Catalunya,
Edifici Nexus, Gran Capit\`a 2,
E-08034 Barcelona, Spain}
\author{P. von Neumann-Cosel}
\affiliation{Institut f\"ur Kernphysik, Technische Universit\"at Darmstadt,
64289 Darmstadt, Germany}
\author{A. Richter}
\affiliation{Institut f\"ur Kernphysik, Technische Universit\"at Darmstadt,
64289 Darmstadt, Germany}

\date{\today}

\begin{abstract}
Highly precise data on the magnetic dipole strength distributions
from the Darmstadt electron linear accelerator for the nuclei
$^{50}$Ti, $^{52}$Cr and $^{54}$Fe are dominated by isovector
Gamow-Teller-like contributions and can therefore be translated
into inelastic total and differential neutral-current neutrino-nucleus cross
sections at supernova neutrino energies. The results agree well
with large-scale shell-model calculations, validating this model.
\end{abstract}

\maketitle

Knowledge about inelastic neutrino-nucleus scattering plays an
important role in many astrophysical applications, including r-process
nucleosynthesis, the synthesis of certain elements like $^{10,11}$B
and $^{19}$F during a supernova explosion by the $\nu$-process or for
the detection of supernova neutrinos (e.g.\ 
see~\cite{Avignone.Chatterjee.ea:2003}).  Although inelastic
neutrino-nucleus scattering is not yet considered in supernova
simulations, several model studies have indicated that it might be
relevant to several aspects of supernova physics $i$) for the neutrino
opacities and thermalization during the collapse phase,
\cite{Bruenn.Haxton:1991}: $ii$) for the revival of the stalled shock
in the delayed explosion mechanism
\cite{Haxton:1988,Balantekin.Fuller:2003} and $iii$) for explosive
nucleosynthesis \cite{Hix.Mezzacappa.ea:2003}.  To predict the outcome
of supernova simulations with confidence a better handle on
neutrino-nucleus interactions is called
for~\cite{Balantekin.Fuller:2003}, in particular on nuclei in the iron
mass range $A \sim 56$~\cite{Hix.Mezzacappa.ea:2003}. While
charged-current neutrino-nucleus reactions -- the inverse of electron
and positron captures -- are included in supernova simulations
\cite{Bruenn:1985}, inelastic neutrino-nucleus scattering is not.
Unfortunately no data for inelastic neutrino-nucleus scattering is
currently available (except for the ground state transition to the
$T=1$ state at 15.11~MeV excitation energy in
$^{12}$C~\cite{Zeitnitz:1994,Auerbach.Burman.ea:2001}).  To measure
some relevant neutrino-nucleus cross sections (mainly in the iron mass
range) a dedicated detector at the Oak Ridge spallation neutron source
has been proposed \cite{Avignone.Chatterjee.ea:2003}.  To sharpen the
experimental program at this facility and to improve supernova
simulations, inelastic neutrino-nucleus cross sections should be
incorporated into the supernova models.  It appears as if the needed
inelastic neutrino cross sections for iron-region nuclei have to be
evaluated by theoretical models without constraint by data. This
manuscript will demonstrate that this is in fact not the case. Our aim
is to show that precision data on the magnetic dipole ($M1$) strength
distributions, obtained by inelastic electron scattering, supply to a
large extent the required information about the nuclear Gamow-Teller
(GT) distribution which determines the inelastic neutrino-nucleus
cross sections for supernova neutrino energies. This intimate relation
of $M1$ and $GT$ strength has already been exploited before to
estimate neutrino cross sections for either individual transitions
(e.g.\ in $^{12}$C \cite{Haxton:1978,Donnelly.Peccei:1979}) or total
cross sections (e.g.\ in $^{208}$Pb
\cite{Fuller.Haxton.Mclaughlin:1999,Jachowicz.Heyde.Ryckebusch:2002}).
We will add to this by demonstrating that large-scale shell-model
calculations agree quite well with the precision $M1$ data, thus
validating the use of such models to determine the required cross
sections for nuclei where no data exist, or at the finite-temperature
conditions in a supernova.

The  $M1$ response is one of the fundamental low-energy
excitations of the nucleus. It can be well explored by means of
inelastic electron scattering. Such transitions are mediated by
the operator
\begin {equation}
  \label{eq:bm1}
  \bm{O}(M1) = \sqrt{\frac{3}{4\pi}} \sum_k [ g_l(k) \bm{l}(k) + g_s(k)
  \bm{s}(k)] \mu_N
\end{equation}
where $\bm{l}$ and $\bm{s}$ are the orbital and spin angular momentum
operators, and the sum runs over all nucleons.  The orbital and spin
gyromagnetic factors are given by $g_l=1$, $g_s= 5.586$ for protons
and $g_l=0$, $g_s=-3.826$ for neutrons \cite{Mohr.Taylor:2000};
$\mu_N$ is the nuclear magneton.  Using isospin quantum numbers $\pm
1/2$ for protons and neutrons, respectively, and
$\bm{t}_0=\bm{\tau}_0/2$; Eq.~(\ref{eq:bm1}) can be rewritten in
isovector and isoscalar parts. Due to a strong cancellation of the
$g$-factors in the isoscalar part, the isovector part dominates.  The
respective isovector $M1$ operator is given by
\begin{equation}
\bm{O}(M1)_{\mathrm{iv}} = \sqrt{\frac{3}{4\pi}} \sum_k [
\bm{l}(k)\bm{t}_0(k) + (g_s^p-g_s^n)\bm{s}(k)\bm{t}_0(k) ] \mu_N.
\end{equation}
We note that the spin part of the isovector $M1$ operator is the
zero component of the $GT$ operator,
\begin{equation}
\bm{O}(GT_0) = \sum_k \bm{\sigma} (k)\bm{t}_0(k) = \sum_k 2
\bm{s}(k)\bm{t}_0(k),
\end{equation}
however, enhanced by the factor $\sqrt{3/4\pi}(g_s^p -
g_s^n) \mu_N/2=2.2993\mu_N$.  On the other hand, inelastic neutrino-nucleus
scattering at low energies, where finite momentum transfer corrections
can be neglected, is dominated by allowed transitions. The cross
section for a transition from an initial nuclear state ($i$) to a
final state ($f$) is given by
\cite{Donnelly.Peccei:1979}
\begin{equation}
\sigma_{i,f} (E_\nu) = \frac{G_F^2 g^2_A}{\pi(2J_i+1)} (E_\nu-\omega)^2
|\langle f||
\sum_k \bm{\sigma}(k) \bm{t}(k) || i \rangle |^2,
\end{equation}
where $G_F$ and $g_A$ are the Fermi and axialvector coupling
constants, respectively, $E_\nu$ is the energy of the scattered
neutrino and $\omega$ is the difference between final and initial
nuclear energies. Note that for ground state transitions $E_x=\omega$.
The nuclear dependence is contained in the $B(GT_0)=g_A^2 |\langle f
|| \sum_k \bm{\sigma}(k) \bm{t}(k) || i \rangle |^2/(2J_i+1)$
reduced transition probability between the initial and final nuclear
states.

Thus, experimental $M1$ data yield the desired $GT_0$ information,
required to determine inelastic neutrino scattering on nuclei at
supernova energies, to the extent that the isoscalar and orbital
pieces present in the $M1$ operator can be neglected.  On general
grounds one expects that the isovector component
dominates over the isoscalar piece. Furthermore, it is well known that
the major strength of the orbital and spin $M1$ responses are
energetically well separated in the nucleus. In $pf$-shell nuclei,
which are of interest for supernova neutrino-nucleus scattering,
the orbital strength is located at excitation energies $E_x
\simeq 2$--4~MeV \cite{Guhr.Diesener.ea:1990}, while the spin $M1$
strength is concentrated between 7 and 11 MeV.  A separation of
spin and orbital pieces is further facilitated by the fact that
the orbital part is strongly related to nuclear
deformation \cite{Enders.Kaiser.ea:1999}. For
example, the  scissors mode \cite{Bohle.Richter.ea:1984},
which is the collective orbital $M1$ excitation, has been detected
in well-deformed nuclei like
$^{56}$Fe~\cite{Fearick.Hartung.ea:2003}. Thus one can expect that
in spherical nuclei the orbital $M1$ response is not only
energetically well separated from the spin part, but also strongly
suppressed.

Examples of spherical $pf$-shell nuclei are $^{50}$Ti, $^{52}$Cr and
$^{54}$Fe. As these nuclei have also the advantage that precise M1
response data exist from high-resolution inelastic electron scattering
experiments \cite{Sober.Metsch.ea:1985} we have chosen these 3 nuclei
for our further investigation. Our strategy now is to show, in a
detailed comparison of data and shell model calculations, that the M1
data indeed represent the desired $GT_0$ information in a sufficient
approximation to transform them into total and differential
neutrino-nucleus cross sections.  All the total strengths and the
strength functions of $^{50}$Ti have been computed using the code
NATHAN~\cite{Caurier.Martinez-Pinedo.ea:1999b}, and the KB3G residual
interaction~\cite{Poves.Sanchez-Solano.ea:2001} in the complete $pf$
model space (orbits $f_{7/2}, p_{3/2},p_{1/2}$, and $f_{5/2}$). For
$^{52}$Cr and $^{54}$Fe the strength functions are computed in
truncated model spaces, allowing up to 6 and 5 protons and neutrons to
be promoted from the lowest $f_{7/2}$ orbital into the other
$pf$-shell orbitals, respectively.  The $M1$ and $GT_0$ response
functions are calculated with 400 Lanczos iterations for both isospin
channels. As customary in shell-model calculations, the spin operator
is replaced by an effective operator $\bm{s}_{\mathrm{eff}} = 0.75
\bm{s}$, where the constant is universal for all $pf$-shell
nuclei~\cite{VonNeumann.Poves.ea:1998}.

\begin{figure}[htb]
  \centering
  \includegraphics[width=\columnwidth]{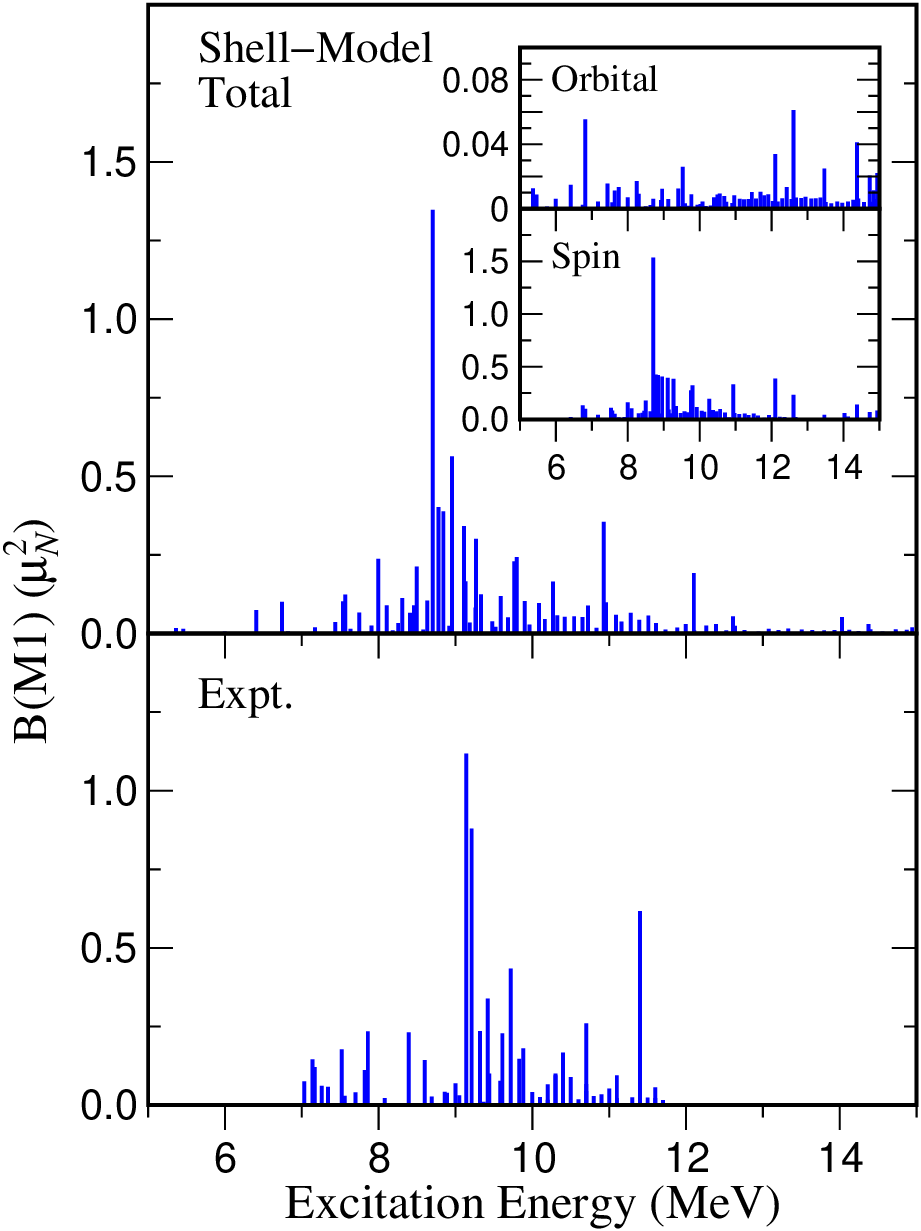}
    \caption{Comparison of experimental $M1$ strength distribution
      [$B(M1)= |\langle f || \bm{O}(M1) || i \rangle|^2/(2J_i+1)$] in
      $^{52}$Cr (bottom) with the shell-model result (top). The inset
      shows the decomposition into spin (botton) and orbital (top)
      parts.  Note the different scales of the ordinate for the spin
      and orbital pieces, respectively.\label{fig:cr52m1}}
\end{figure}

Experimentally $M1$ data have been determined for the energy
intervals 8.5--11.6~MeV in $^{50}$Ti (resolving the $M1$ strength
for 29 individual states), while for the other two nuclei $M1$
data exist for the energy interval 7.0--11.8~MeV resolving 53
states for $^{52}$Cr and 33 states for $^{54}$Fe. The summed
experimental $B(M1)$ strength (in $\mu_N^2$) in these intervals is
4.5(5) for $^{50}$Ti, 8.1(5) for $^{52}$Cr and 6.6(4) for
$^{54}$Fe, which for $^{50}$Ti and $^{52}$Cr, is in agreement with
the shell model ($4.3$ and $7.6$, respectively, in the same
intervals). For $^{54}$Fe the shell model strength is slightly
larger ($8.6$) than the data, which is also true, if another
interaction (GXPF1 \cite{Honma.Otsuka.ea:2004}) is used ($8.4$). The total
shell model $B(M1)$ strengths of 7.2 for $^{50}$Ti, 8.7 for
$^{52}$Cr and 10.2 for $^{54}$Fe indicate some additional strength
outside of the experimental energy window.
For a comparison of the $M1$ strength distributions a problem
arises due to uncertainties of the distinction between $M1$ and
$M2$ transitions in some of the ($e,e'$) data. Therefore, all
possible $M1$ candidates are modified by the weighing factors
introduced in \cite{Sober.Metsch.ea:1985} to express the level of
confidence of the assignment. The experimental sensitivity limit
$B(M1) \simeq 0.04 \mu_N^2$ is also taken into account for
comparison with the model results. It should be noted that, where
data are available \cite{Berg.Rueck.ea:1981,Wesselborg:1993} good
agreement with nuclear resonance fluorescence experiments is
observed for the prominent $M1$ transitions. This is also the case
for other $pf$-shell nuclei
\cite{Degener.Blasing.ea:1990,Bauwens.Bryssinck.ea:2000}. For all
nuclei, the energy dependence of the observed $M1$ strength
distribution is well reproduced. This is shown in
Fig.~\ref{fig:cr52m1} for the example of $^{52}$Cr.

To determine how well the $M1$ data might reflect the desired
$GT_0$ information we have performed shell-model calculations for
the individual orbital and spin parts of the $M1$ operator as well
as calculations for the $GT_0$ operator, which, except for a
constant factor, represents the isovector spin contribution to the
$M1$ operator.  The results are displayed in
Fig.~\ref{fig:cr52m1}. As expected for spherical nuclei,
the orbital $M1$ strength is significantly smaller,
by about an order of magnitude, than the spin $M1$ strength.  The
interference between the orbital and spin parts is state-dependent
and is largely cancelled out, when the strength is averaged over
several states. A similar situation occurs for the isoscalar spin
contribution, but now its contribution to the total strength is
even smaller.

\begin{figure}[htb]
  \centering
  \includegraphics[width=1.0\columnwidth]{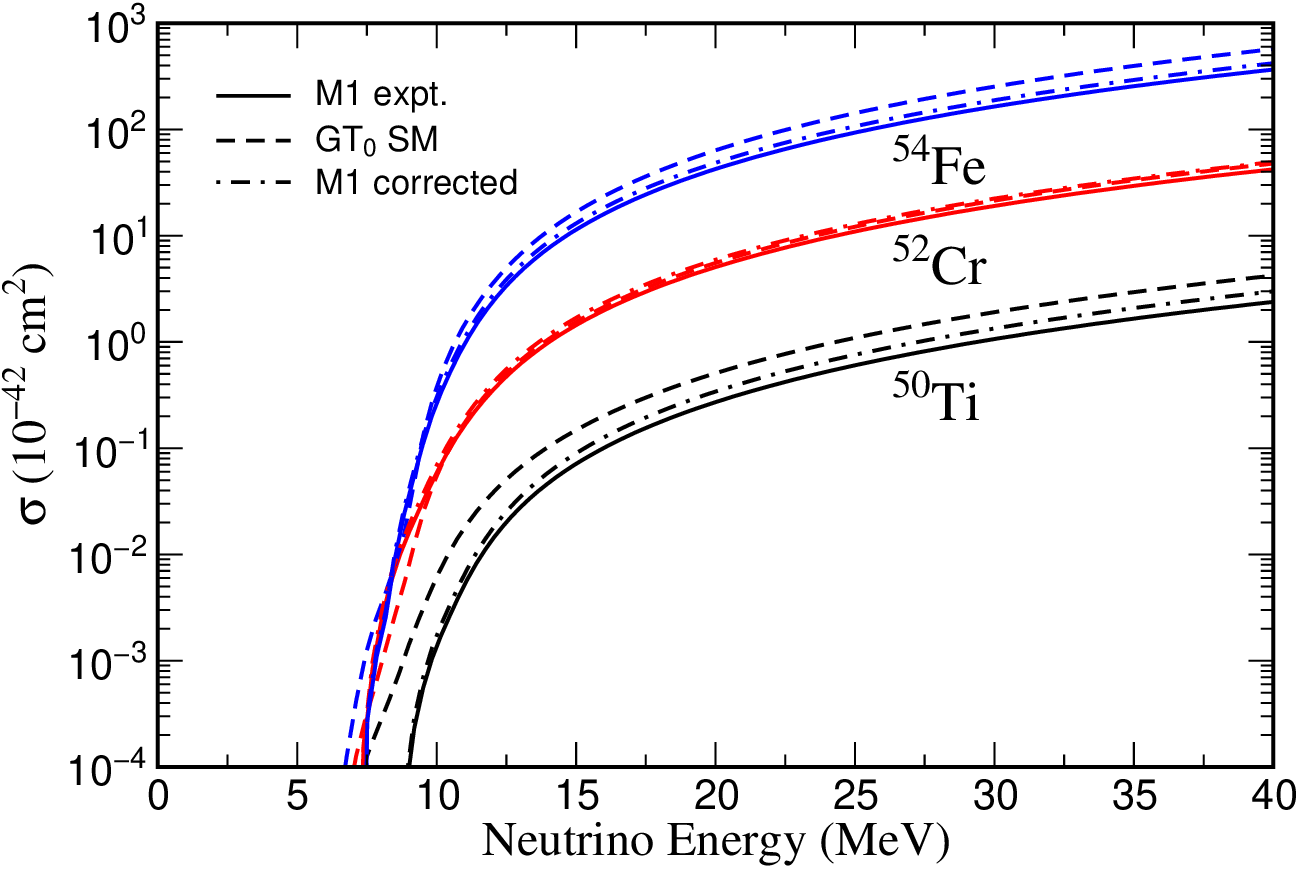}
  \caption{Neutrino-nucleus cross sections, calculated from the M1
    data (solid lines) and the shell-model $GT_0$ distributions
    (dotted) for $^{50}$Ti (multiplied by 0.1), $^{52}$Cr, and
    $^{54}$Fe (times 10). The long-dashed lines show the  cross
    sections from the $M1$ data,
    corrected for possible strength outside the experimental
    energy window. \label{fig:M1vsGT0}}
\end{figure}

Supernova simulations require differential neutrino-nucleus cross
sections as functions of initial and final neutrino energies, where
neutrinos of different flavors are comprised in energy bins of a few
MeV~\cite{Liebendoerfer.Messer.ea:2004,Rampp.Janka:2002,Burrows.Young.ea:2000},
i.e., cross sections are averaged over many final nuclear states.
Cancelling most of the interference between orbital and spin
contributions, the $M1$ data should represent the desired $GT_0$
information, simply using the relation $B(M1) = 3(g_s^p-
g_s^n)^2\mu_N^2/(16g^2_A\pi) B(GT_0)$.  Figure~\ref{fig:M1vsGT0}
compares the total neutrino-nucleus cross sections for the 3 nuclei,
calculated from the experimental $M1$ data with those obtained from
the shell-model $GT_0$ distribution.  As some of the $M1$ strength is
predicted to reside outside of the currently explored experimental
energy window,
we have corrected for this by multiplying
the ``$M1$ cross section'' with the ratio
$B(GT_0)/B(GT_0, \Delta E)$, where $\Delta E$
defines the experimental energy interval and the ratio is taken
from the shell-model calculations.
Based on the above theoretical discussion one can assume that the
(energetically complete) ``$M1$ cross section'' represents the
neutrino-nucleus cross sections quite well.

\begin{figure}[htb]
  \centering
  \includegraphics[width=1.0\columnwidth]{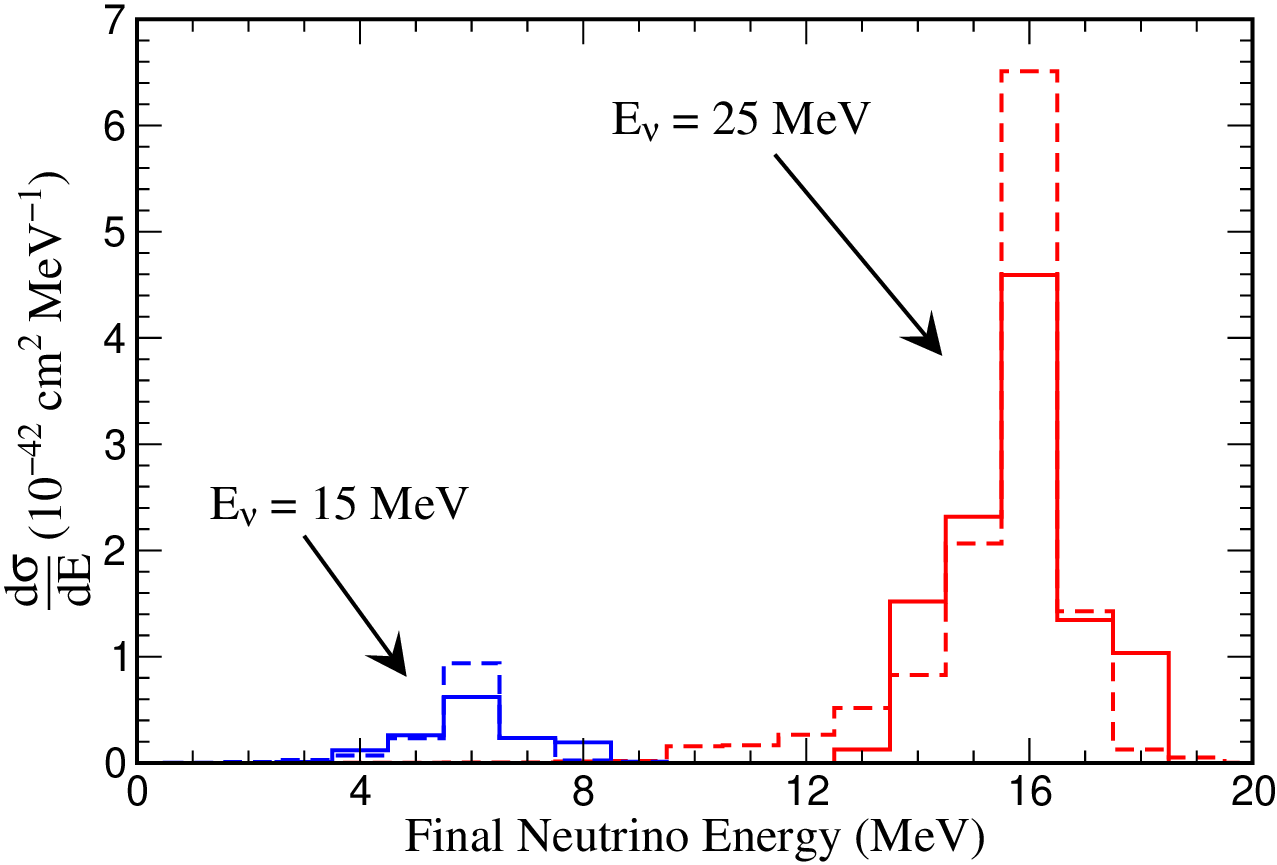}
  \caption{Differential inelastic neutrino cross sections
    for $^{52}$Cr and initial neutrino energies $E_\nu=15$ MeV and 25
    MeV.  The solid histograms are obtained from the $M1$ data, the
    dashed from shell-model calculations.  The final neutrino energies
    are given by $E_f=E_\nu-\omega$. \label{fig:doubl}}
\end{figure}

Figure~\ref{fig:doubl} shows the differential neutrino cross section
for $^{52}$Cr at two representative supernova neutrino energies. The
cross sections, obtained from the experimental M1 data and the shell
model, agree quite well, if binned in energy intervals of a resolution
(1~MeV or somewhat larger) as required in supernova simulations.

\begin{figure}[htb]
  \centering
  \includegraphics[width=1.0\columnwidth]{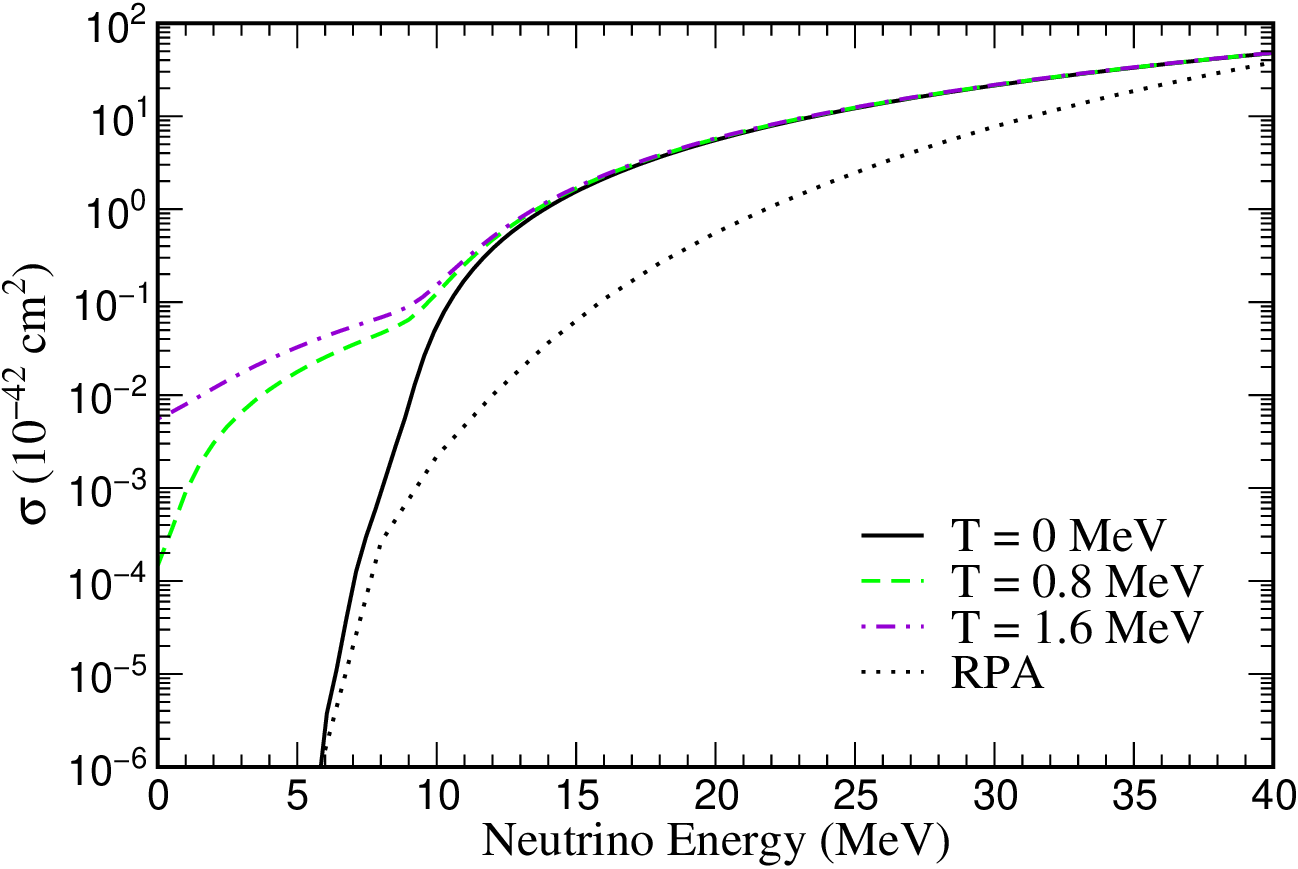}
    \caption{Inelastic neutrino scattering cross section on $^{52}$Cr,
      calculated on the basis of shell-model $GT_0$ distributions at
      finite temperatures. The dotted curve represents the RPA
      contributions of other multipoles to the cross sections,
      including finite-momentum transfer corrections.
      \label{fig:temp}}
\end{figure}

The comparison of $M1$ and theoretical cross sections suggests that
shell-model based calculations of inelastic neutrino scattering at
supernova relevant energies are quite accurate and hence the shell
model is the method of choice to determine the cross section for the
many nuclei in the iron mass region needed in core-collapse
simulations. However, such cross sections require additional
considerations so far neglected. These must include the effects of
finite momentum transfer, of the finite temperature in the supernova
environment and the contributions of additional (forbidden) multipoles
to the cross section.  The latter become only relevant for neutrino
energies which are sufficiently larger than the centroid energy of the
respective giant resonance of this multipole. At such neutrino
energies the cross section depends only on the total strength of the
multipole and its approximate centroid energy (and not on a detailed
reproduction of the strength distribution) and is well described
within the Random Phase Approximation (RPA)
\cite{Kolbe.Langanke.ea:2003}. We have calculated the RPA contribution
to the cross section arising from multipoles other than $GT_0$, using
the formalism of
\cite{Kolbe.Langanke.Vogel:1999,Kolbe.Langanke.ea:2003} which
explicitly considers the finite-momentum dependence of the multipole
operators.  For the $GT_0$ component the finite momentum transfer
corrections can be considered as described in
\cite{Toivanen.Kolbe.ea:2001}.  Following the approach of
\cite{Sampaio.Langanke.ea:2002} we have derived the finite-temperature
corrections to the cross sections from the shell model $GT_0$
transitions between a few hundred excited states and the 6 lowest
nuclear states.  The $^{52}$Cr cross sections are presented in
Fig.~\ref{fig:temp}. Due to the thermal population of excited initial
states the neutrino cross sections are significantly enhanced at low
energies during the early collapse phase ($E_\nu \sim 10$ MeV).  Once
the neutrino energy is large enough to allow scattering to the
centroid of the $GT_0$ strength, which resides at energies around
8--11~MeV, finite temperature effects become unimportant and the
neutrino cross section can be derived effectively from the ground
state distribution, as discussed in \cite{Sampaio.Langanke.ea:2002},
and thus is directly constrained by the $M1$ data. This applies to the
neutrino energy regime relevant to post-shock supernova simulations.
Contributions from multipoles other than the $GT_0$ become important
for $E_\nu >20$ MeV and dominate for energies higher than 35 MeV.

In summary, we have translated the high-precise $(e,e')$ M1 data for
$^{50}$Ti,$^{52}$Cr, and $^{54}$Fe, into detailed total and
differential inelastic neutral-current neutrino-nucleus cross sections. Besides
representing for the first time detailed neutral-current cross
sections for nuclei, such data are in particular important for
supernova simulations as they allow to constraint theoretical models
needed to derive the inelastic neutrino-induced cross sections for the
many nuclei in the medium-mass range present in a supernova
environment.
We have further demonstrated that large-scale shell model calculations
are able to describe the data, even in details. Following this
validation, shell model calculations for inelastic neutrino cross
sections on supernova-relevant nuclei are now in progress.

\begin{acknowledgments}
KL is partly supported by the Danish Research Council. GMP is
supported by the Spanish MCyT and by the European Union ERDF under
contracts AYA2002-04094-C03-02 and AYA2003-06128. PvNC and AR
acknowledge support by the DFG under contract SFB 634. Computational
cycles were provided by the Centre for Scientific Computing in
{\AA}rhus.
\end{acknowledgments}

\end{document}